\documentclass[aps,prl,twocolumn,superscriptaddress]{revtex4-2}

\usepackage{mathrsfs}
\usepackage{amsmath}
\usepackage{amssymb}
\usepackage{bm}
\usepackage{graphicx}
    \graphicspath{{Figures/}}
\usepackage{hyperref}
\usepackage{physics}
\usepackage{braket}
\usepackage{xcolor}
\usepackage{siunitx}
\usepackage[caption=false]{subfig}
\usepackage{dsfont}
\usepackage[version=4]{mhchem}

\definecolor{darkgreen}{rgb}{0.0, 0.5, 0.0}

\hypersetup{
    colorlinks=true,
    linkcolor=blue,
    citecolor=blue,
    urlcolor=blue
}

\begin{document}

\title{Hidden Frustration in Collinear Altermagnets: 
\\
Pairing Vortices and Equilibrium Spin Current Loops}

\author{Vincent P. Flynn}
\email{flynnvin@bc.edu}
\affiliation{Department of Physics, Boston College, Chestnut Hill, Massachusetts 02467, USA}

\affiliation{Department of Physics and Astronomy, Dartmouth College, Hanover, New Hampshire 03755, USA}

\author{Benedetta Flebus}
\affiliation{Department of Physics, Boston College, Chestnut Hill, Massachusetts 02467, USA}

\date{\today}

\begin{abstract}
We show that a magnet can remain perfectly collinear while its quantum vacuum circulates. In a centrosymmetric altermagnet, a symmetry-allowed locally staggered Dzyaloshinskii–Moriya coupling imprints a gauge-irremovable vortex-antivortex pair into the anomalous magnon pair correlations at high-symmetry points in momentum space. In real space, these hidden vortices produce an antiferrochiral array of equilibrium spin currents circulating oppositely around neighboring plaquettes. Gauge-irremovable frustration therefore survives despite classical collinearity: it exists entirely in the quantum correlations. Our results show that complex pairing, gauge-invariant fluxes, and equilibrium loop currents—structures encountered across electronic flux phases, spin liquids, and frustrated quantum magnets—can be encoded in the squeezed vacuum of a collinear magnet.
\end{abstract}

\maketitle

{\em Introduction.---}  Frustration and hidden circulation are recurring motifs in attempts to understand the most elusive phases of quantum matter. Anderson's proposal that a quantum antiferromagnet (AFM) might evade N\'eel order through resonating singlets \cite{Anderson1973} later became central to theories of high-temperature superconductivity, where doping reorganizes the correlations of a Mott AFM into competing states with no simple classical counterpart \cite{Anderson1987,ShraimanSiggia1989}.
Geometrically frustrated magnets offer another setting, with competing exchange constraints built directly into the lattice: on the kagome lattice quantum fluctuations melt magnetic order into a spin liquid, while on the triangular lattice they renormalize the classical $120^\circ$ state and, in a magnetic field, stabilize the collinear up-up-down plateau or a condensate of bound magnon pairs \cite{YanHuseWhite2011,Capriotti1999,AliceaChubukovStarykh2009,ChubukovStarykh2013}. In these familiar cases, quantum fluctuations determine whether and how the moments order. A subtler possibility is that the static spin configuration remains deceptively simple while the correlations between spins acquire a nontrivial phase structure.

Hidden circulation is one manifestation of such structures. A canonical example is the family of flux phases proposed for doped Mott insulators: in the Affleck--Marston construction, complex spinon bond amplitudes generate an emergent $\pi$ flux~\cite{AffleckMarston1988,MarstonAffleck1989}, predicted to manifest as charge currents circulating in opposite directions around neighboring plaquettes~\cite{HsuMarstonAffleck1991}, while the $d$-density-wave proposal realizes the same staggered-current motif through an imaginary particle--hole condensate with a $d$-wave form factor~\cite{ChakravartyLaughlinMorrNayak2001}.
A magnetic counterpart was proposed near the endpoint of the one-third plateau of an anisotropic triangular AFM, where the condensation of bound pairs of soft magnons yields vector-chiral order and alternating equilibrium spin currents, without transverse dipolar order \cite{ChubukovStarykh2013}. While these theories are microscopically distinct, they share a common structure: a complex bond or pair amplitude whose phase encodes gauge-invariant information around closed lattice paths. 

\begin{figure}[t!]
    \centering
\includegraphics[width=.8\columnwidth]{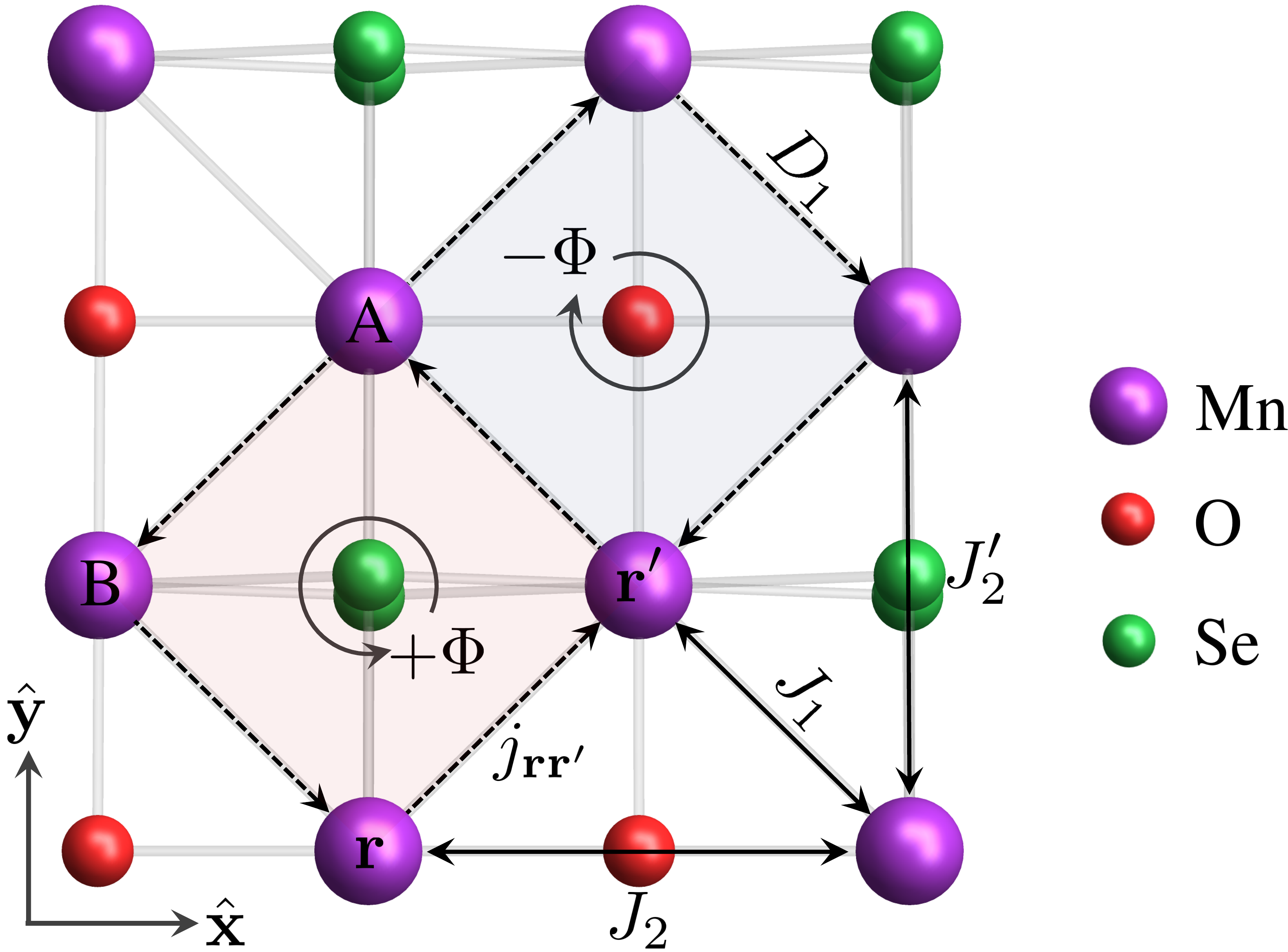}
    \caption{An \ce{MnSeO} layer of the $d$-wave altermagnet LOMS,
shown as an illustrative realization of the model under consideration.  The nearest and next-nearest neighbor symmetric exchanges $J_1$ and $(J_2,J_2')$, respectively, are shown with double-sided arrows. Nearest neighbors are coupled via a DMI of strength $D_1$ along $\hat{\mathbf{z}}$. An arrow pointing from $\mathbf{S}_1$ to $\mathbf{S}_2$ represents a term $+D_1\hat{\mathbf{z}}\cdot (\mathbf{S}_1\times \mathbf{S}_2)$ in the Hamiltonian Eq.\,\eqref{eq: Ham}. Staggered enclosed fluxes $\pm\Phi$ responsible for chiral equilibrium spin currents $j_{\mathbf{r}\mathbf{r}'}$ in Eq.\,\eqref{eq: jrrp} are shown in neighboring plaquettes.}
    \label{fig: lattice}
\end{figure}

In this Letter, we show that the same ingredients arise in a strikingly simpler setting: the zero-point fluctuations of a collinear, centrosymmetric altermagnetic insulator \cite{vsmejkal2020SciAdv,vsmejkal2022AMPRX,vsmejkal2022emergingPRX}. We consider the inverse Lieb lattice shown in Fig. \ref{fig: lattice} realized, for instance, in the $d$-wave
altermagnet \ce{La2O3Mn2Se2} (LOMS) \cite{LOMSNi2010,LOMSKoo2012,LOMSKen2025,LOMSAsai2026,LOMSGarciaGassull2026,LOMSzhang2026}. 
A staggered out-of-plane Dzyaloshinskii–Moriya interaction, allowed by local inversion breaking, induces complex two-magnon squeezing that vanishes at the isolated high-symmetry points $X$ and $Y$ - points that can further host dispersion valleys, forming a doublet reminiscent of the incipient $X/Y$ stripe fields of spin-driven nematic theories \cite{XuMullerSachdev2008}. Consequently, the phase of the anomalous magnon pair correlations
winds oppositely around the two zeros and cannot be removed by any smooth, globally consistent choice of transverse-spin frame. This structure additionally results in a nontrivial quadrupolar Berry curvature. In real space, these pairing vortices manifest as opposing gauge-irremovable pairing fluxes through neighboring plaquettes, producing an antiferrochiral texture of counter-circulating equilibrium spin currents. A centrosymmetric altermagnet whose classical order is entirely collinear thus hosts a frustrated, current-carrying quantum vacuum.

{\em Model.---} We consider the 2D inverse Lieb lattice shown in Fig.\,\ref{fig: lattice}. Spin-$S$ operators
$\mathbf{S}_{\mathbf{r}}^\alpha$ reside on two magnetic
sublattices $\alpha=A,B$, whose sites are $
\mathbf{r}\in\Lambda^{\alpha}
=
\left\{
(n,m)a+\bm{\delta}^{\alpha}:(n,m)\in\mathbb Z^2
\right\},$ where $a>0$ is the lattice constant, $\bm{\delta}^{A}=(a/2,0)$, and $\bm{\delta}^{B}=(0,a/2)$. Two inequivalent families of nonmagnetic mediating atoms occupy the sites $(n,m)a$ and $(n+1/2,m+1/2)a$, respectively. Henceforth, we set $a=1$. 

The minimal spin Hamiltonian $H= H_A + H_B + H_{AB}$ is given by
\begin{align}
H_A &= \sum_{\mathbf{r}\in \Lambda^A} \left( J_2 \mathbf{S}^A_\mathbf{r}\cdot \mathbf{S}^A_{\mathbf{r} +\hat{\mathbf{x}}}+J_2' \mathbf{S}^A_\mathbf{r}\cdot \mathbf{S}^A_{\mathbf{r} +\hat{\mathbf{y}}} -K (S^A_\mathbf{r}{}^z)^2\right) \nonumber 
\\
H_B &= \sum_{\mathbf{r}\in \Lambda^B} \left( J_2' \mathbf{S}^B_\mathbf{r}\cdot \mathbf{S}^B_{\mathbf{r} +\hat{\mathbf{x}}}+J_2 \mathbf{S}^B_\mathbf{r}\cdot \mathbf{S}^B_{\mathbf{r} +\hat{\mathbf{y}}} -K (S^B_\mathbf{r}{}^z)^2\right) \nonumber
\\
H_{AB} &= \sum_{\mathbf{r}\in \Lambda^A}\sum_{\bm{\delta}}\left[ J_1 \mathbf{S}_\mathbf{r}\cdot \mathbf{S}_{\mathbf{r}+\bm{\delta}}  +D_{\bm{\delta}} \hat{\mathbf{z}}\cdot( \mathbf{S}_\mathbf{r}\times \mathbf{S}_{\mathbf{r}+\bm{\delta}}) 
\right]  \label{eq: Ham}
\end{align}
where the sum over $\bm{\delta}$ is over the four diagonal bonds $\pm \bm{\delta}^\sigma = \pm(1/2,\sigma /2)$, $\sigma\in\{\pm\}$. Here $J_1>0$ couples the two magnetic sublattices antiferromagnetically,
whereas $K>0$ sets the collinear easy-axis. The inequivalence
$J_2\neq J_2'$ between the two intra-sublattice exchange paths produces the $d$-wave altermagnetic splitting. The absence of inversion symmetry at the nearest-neighbor $A$-$B$ bond centers permits an out-of-plane Dzyaloshinskii--Moriya interaction (DMI), whose sign alternates between the two diagonal bond orientations: $D_{\bm{\delta}}=\pm D_1$ for $\bm{\delta}\parallel\bm{\delta}^{\pm}$ \cite{syljuasen2025quantum}. This system also has global U(1) symmetry $\mathbf{S}_{\mathbf{r}}\mapsto R_z(\varphi)\mathbf{S}_{\mathbf{r}}$ with $R_z(\varphi)$ a rotation by $\varphi$ about $\hat{\mathbf{z}}$.

We focus on the parameter regime in which Eq.\,\eqref{eq: Ham} supports a collinear G-type AFM ground state, with the two magnetic sublattices polarized antiparallel along $\hat{\mathbf{z}}$: $\mathbf{S}_\mathbf{r} = S\hat{\mathbf{z}}$ for $\mathbf{r}\in \Lambda^A$ and $-S\hat{\mathbf{z}}$ for $\mathbf{r}\in\Lambda^B$. This state is stabilized by a dominant AFM exchange $J_1$, as realized in the intensely studied altermagnetic candidate LOMS \cite{LOMSNi2010,LOMSKoo2012,LOMSKen2025,LOMSGarciaGassull2026,LOMSzhang2026,LOMSAsai2026}. Although microscopic estimates agree on the dominance of $J_1$, the signs and relative magnitudes of $J_2$ and $J_2'$ remain unsettled. As shown below, either sign of $J_2'$ can preserve the same collinear ground state while producing qualitatively distinct, previously unexplored magnon dispersions.

Previous models of LOMS have omitted the symmetry-allowed DMI in Eq.\,\eqref{eq: Ham}. Although recent microscopic analysis did not determine $D_1$, it showed that the dominant $J_1$ contains an essential ligand-assisted contribution \cite{LOMSGarciaGassull2026}. The nearly quenched orbital moment of Mn$^{2+}$ therefore does not, by itself, justify setting $D_1=0$: the same virtual paths sample inequivalent ligand orbitals and can transmit spin--orbit coupling to the effective Mn--Mn interaction.  We thus retain $D_1$ as a symmetry-allowed coupling. Crucially, the phenomena uncovered in this Letter require only $J_1,D_1 \neq 0$, and therefore arise generically whenever this symmetry-allowed interaction is present.

{\em Altermagnetic dispersion. ---}
We perform a standard Holstein-Primakoff expansion of Eq.\,\eqref{eq: Ham} around the collinear G-type AFM state, followed by a Fourier transform to obtain the quadratic magnon Hamiltonian
\begin{align}\label{eq: magHam}
    H = S\sum_{\mathbf{k}}\Phi_\mathbf{k}^\dag\mathcal{H}_\mathbf{k}\Phi_\mathbf{k},\quad \mathcal{H}(\mathbf{k}) = \begin{pmatrix}
        \varepsilon^A_\mathbf{k} & \Delta_\mathbf{k} \\ \Delta_\mathbf{k}^* & \varepsilon^B_{-\mathbf{k}}
    \end{pmatrix} 
\end{align}
where $\Phi_\mathbf{k} =(a_\mathbf{k},b_{-\mathbf{k}}^\dag)^T$ is the magnon Nambu array, $\varepsilon^A_\mathbf{k} = J+2J_2\cos(k_x)+2J_2' \cos(k_y)$, $\varepsilon^B_{-\mathbf{k}} = J+2J_2'\cos(k_x)+2J_2 \cos(k_y)$, $J=2(2J_1 -J_2-J_2'+K)$, and
\begin{align}\label{eq: Deltak}
 \Delta_{\mathbf{k}}
=g\left[
e^{i\phi}\cos\left(\mathbf{k}\cdot \bm{\delta}^-\right)+e^{-i\phi}\cos\left(\mathbf{k}\cdot \bm{\delta}^+\right)
\right]
\end{align} 
with $g=2\sqrt{J_1^2+D_1^2}$ and $\tan\phi=D_1/J_1$. 
In what follows, we write $\Delta_{\mathbf{k}} = |\Delta_{\mathbf{k}}| e^{i\phi_\mathbf{k}}$ and note that $\phi_\mathbf{k} = 0$ when $D_1=0$. Both dynamical and thermodynamical stability \cite{Decon} (and thus, stability of the collinear G-type state) are guaranteed by the conditions $\varepsilon^A_\mathbf{k}\varepsilon^B_{-\mathbf{k}}>|\Delta_\mathbf{k}|^2$ and $\varepsilon^{A/B}_\mathbf{k}>0$. These hold for sufficiently large $J_1$.

\begin{figure}[t!]
    \centering
\includegraphics[width=.75\columnwidth]{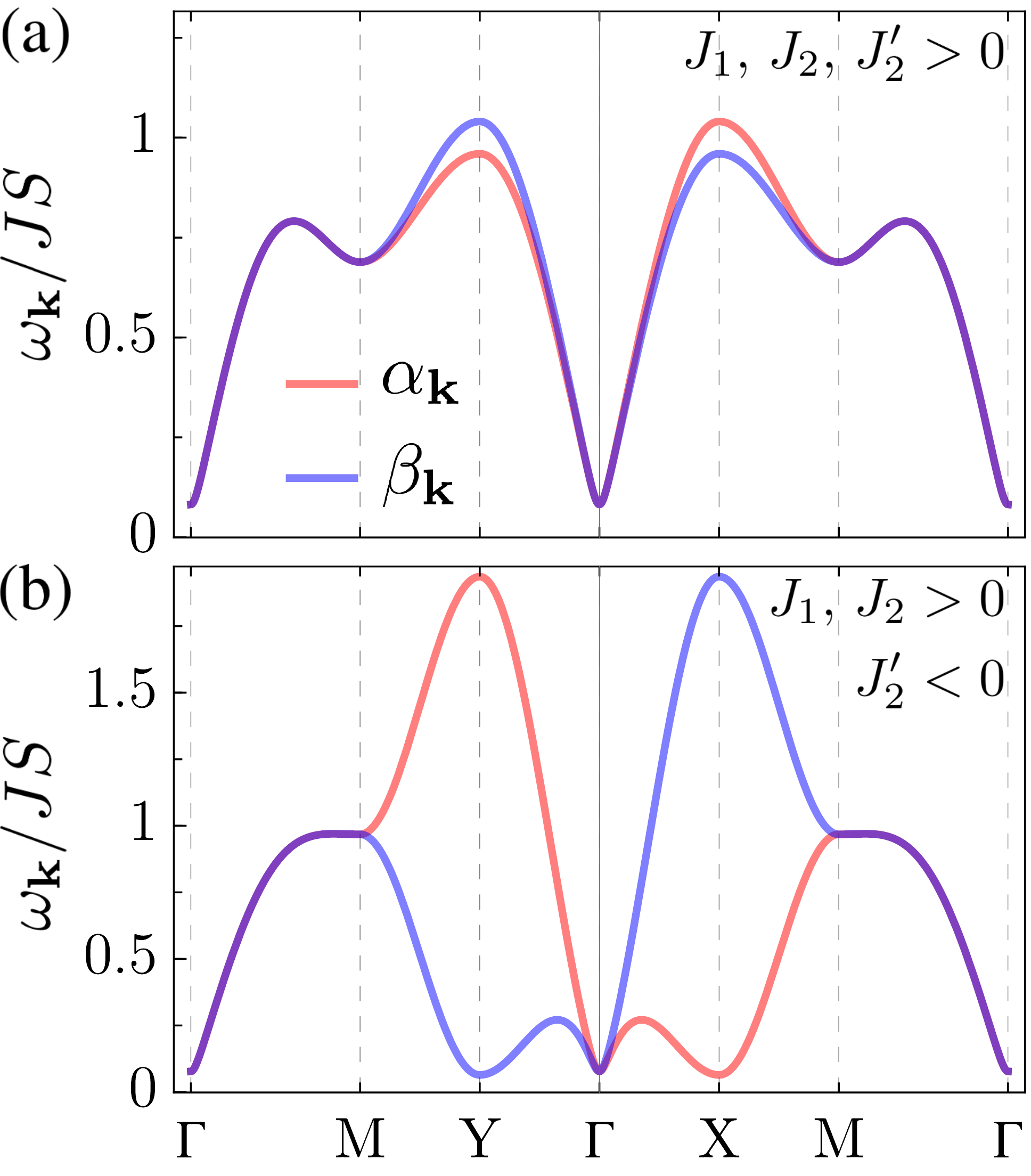}
    \caption{The altermagnetic magnon dispersion Eq.\,\eqref{eq: magdisp}. \textbf{(a)} The case of all AFM couplings using the experimental parameters in Table I of Ref.\,\cite{LOMSAsai2026}. \textbf{(b)} The case with $J_2'=\SI{-1.55}{\milli\electronvolt}$ FM, with $J_1= \SI{1.6}{\milli\electronvolt}$ and $J_2 = \SI{1.5}{\milli\electronvolt}$ both AFM. Here, $X$ and $Y$ are global minima. In both plots, $K=\SI{0.01}{\milli\electronvolt}$ and $D_1 = \SI{0.5}{\milli\electronvolt}$.}
    \label{fig: dispersion}
\end{figure}

We diagonalize Eq.\,\eqref{eq: magHam} via a Bogoliubov transformation
\begin{subequations}\label{eq: Bog}
\begin{align}
    a_\mathbf{k} &= \cosh\left(\frac{\theta_\mathbf{k}}{2}\right)\alpha_\mathbf{k}-e^{i\phi_\mathbf{k}}\sinh\left(\frac{\theta_{\mathbf{k}}}{2}\right)\beta_{-\mathbf{k}}^\dag
    \\
    b_{-\mathbf{k}}^\dag &= \cosh\left(\frac{\theta_\mathbf{k}}{2}\right)\beta_{-\mathbf{k}}^\dag-e^{-i\phi_\mathbf{k}}\sinh\left(\frac{\theta_{\mathbf{k}}}{2}\right)\alpha_{\mathbf{k}},
\end{align}
\end{subequations}
where $\theta_\mathbf{k}$ is the squeezing angle defined via $\tanh \theta_\mathbf{k} = 2|\Delta_\mathbf{k}|/(\varepsilon^A_\mathbf{k} + \varepsilon^B_{-\mathbf{k}})$, and $\alpha_\mathbf{k}$ and $\beta_\mathbf{k}$ are the quasiparticles with dispersions (shown in Fig.\,\ref{fig: dispersion})
\begin{align}\label{eq: magdisp}
\omega^{\alpha/\beta}_\mathbf{\pm\mathbf{k}} = \frac{S}{2}\left[\sqrt{(\varepsilon^A_\mathbf{k}+\varepsilon^B_{-\mathbf{k}})^2-4|\Delta_\mathbf{k}|^2} \pm \left( \varepsilon^A_\mathbf{k}-\varepsilon^B_{-\mathbf{k}} \right)\right]
\end{align}
In Fig.\,\ref{fig: dispersion}, we observe two characteristic features of these dispersions: (i) the bands are spin-split with $d_{x^2-y^2}$ splitting $\propto$ $(J_2-J_2')(\cos(k_x)-\cos(k_y))$; and (ii) both bands are extremal at $\mathbf{k} = X = (\pi,0)$ and $\mathbf{k}=Y=(0,\pi)$, i.e., $\nabla_{\mathbf{k}}\omega^{\alpha/\beta}_{\mathbf{k}}\big|_{\mathbf{k}=X,Y} = 0$. In particular, when 
$|J|J_- > \text{max}(4J_1^2 - JJ_+,4D_1^2+JJ_+)$ with $J_\pm = J_2\pm J_2'$
, $X$ and $Y$  become local minima of the $\alpha$ and $\beta$ branches, respectively - see Fig.\,\ref{fig: dispersion}(b). The two valleys form an altermagnetic doublet exchanged by $C_{4z}\mathcal{T}$. Near softening (as in Fig.\,\ref{fig: dispersion}(b)), this doublet is the bosonic analogue of the incipient $X/Y$ stripe fields in spin-driven nematic theories, with two-magnon squeezing replacing the anomalous electron pairing.

\begin{figure*}
    \centering
    \includegraphics[width=\linewidth]{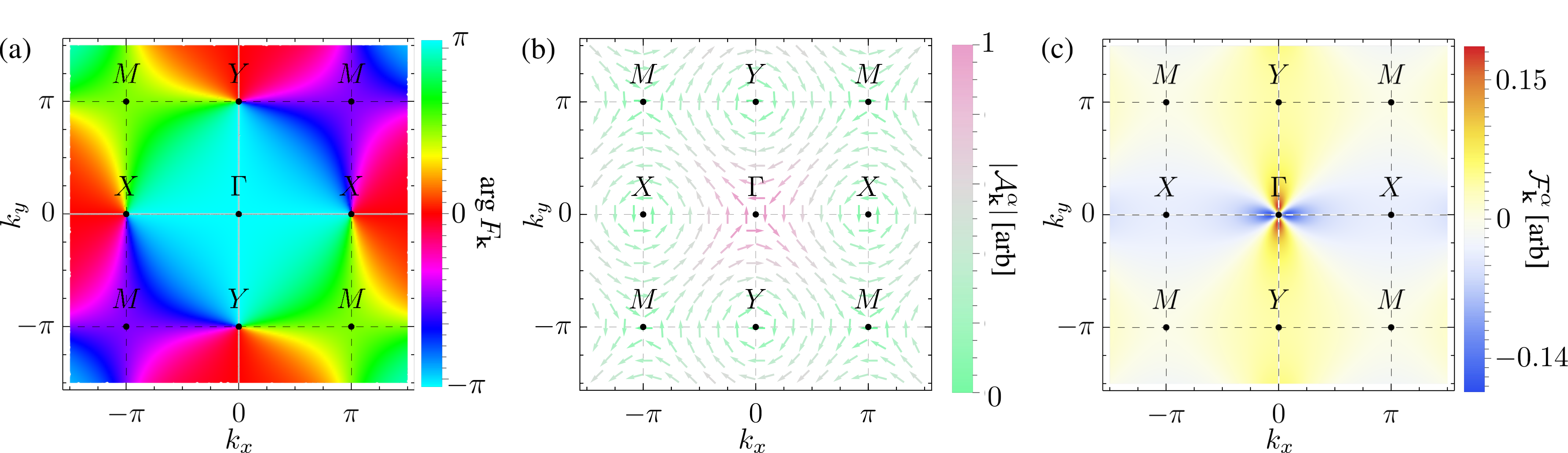}
    \caption{\textbf{(a)} Phase of the anomalous correlator $F_\mathbf{k} = \braket{a_\mathbf{k} b_{-\mathbf{k}}}$. A counterclockwise circuit around $Y$ ($X$) accumulates a phase of $(-)2\pi$. \textbf{(b)} A vector plot of the $\alpha$-mode Berry connection in Eq.\,\eqref{eq: BerryConn}. Oppositely oriented vortices are hosted at $X$ and $Y$. \textbf{(c)} The corresponding Berry curvature $\mathcal{F}^\alpha_\mathbf{k}$. For all three plots, we use the parameters of Fig.\,\ref{fig: dispersion}(b).}
    \label{fig: vortices}
\end{figure*}

{\em Squeezing vortices and chiral spin correlations.---}
For $J_1,D_1\neq 0$, the squeezing function Eq.\,\eqref{eq: Deltak} has two isolated simple zeros: one at $X$ and one at $Y$. Locally,
\begin{subequations}
    \begin{align}
        \Delta_{X+\delta\mathbf{k}} &\approx -2(J_1 \delta k_x - i D_1 \delta k_y),
        \\
        \Delta_{Y+\delta\mathbf{k}} &\approx -2(J_1 \delta k_y - i D_1 \delta k_x),
    \end{align}
\end{subequations}
resulting in opposite winding numbers
\begin{align}
    \nu_X = \frac{1}{2\pi}\oint \nabla_\mathbf{k}\phi_\mathbf{k}\cdot d\mathbf{k} = -1 = -\nu_Y,
\end{align}
where the above contours are positively oriented and chosen to enclose only $X$ or $Y$. Physically, this vortex--antivortex structure imprints itself in the ground state anomalous correlators
$F_{\mathbf{k}}=\langle a_{\mathbf{k}}b_{-\mathbf{k}}\rangle
=-\frac{1}{2}e^{i\phi_{\mathbf{k}}}\sinh\theta_{\mathbf{k}}$.
As $\mathbf{k}$ encircles a zero, $F_{\mathbf{k}}$ executes one complete rotation with opposite orientations around $X$ and $Y$ - see Fig.\,\ref{fig: vortices}(a). The vortex--antivortex pair therefore represents a winding of the ground-state correlations themselves. Additionally, $F_\mathbf{k}$ describes, at quadratic order, collinear transverse correlations through its real part and vector chirality through its imaginary part
\begin{align}\label{eq: Fk}
F_{\mathbf{k}} =  \frac{1}{2S}\langle\big[
 &S^A_{\mathbf{k},x} S^B_{-\mathbf{k},x}+ S^A_{\mathbf{k},y} S^B_{-\mathbf{k},y} \nonumber
 \\
 - i( &S^A_{\mathbf{k},x} S^B_{-\mathbf{k},y}- S^A_{\mathbf{k},y} S^B_{-\mathbf{k},x} ) \big]\rangle.
\end{align}
Vortices thus serve as topological obstructions to unwinding these correlations by means of a smooth local U(1) transformation $F_\mathbf{k}\mapsto e^{i\chi(\mathbf{k})}F_{\mathbf{k}}$. 

Finally, pairing vortices are embedded into eigenmode structure via the pseudo-Hermitian Berry connection \cite{shindou2013topological}
\begin{align}\label{eq: BerryConn}
    \mathcal{A}^\alpha_{\mathbf{k}} = i\braket{\alpha_\mathbf{k}|\sigma_3 \nabla_{\mathbf{k}}|\alpha_{\mathbf{k}}} = -\sinh^2\left(\frac{\theta_{\mathbf{k}}}{2}\right) \nabla_{\mathbf{k}}\phi_\mathbf{k},
\end{align}
and similarly for $\beta$,
where $\ket{\alpha_\mathbf{k}}$ is the eigenvector of the pseudo-Hermitian dynamical matrix $\sigma_3\mathcal{H}_\mathbf{k}$ corresponding to the $\alpha$ eigenmode. The phase gradient in Eq.\,\eqref{eq: BerryConn} is thus sensitive to the squeezing vortices resulting in a quadrupolar Berry curvature $\mathcal{F}^\alpha_\mathbf{k} = \partial_{k_x}\mathcal{A}^{\alpha}_{\mathbf{k},y}-\partial_{k_y}\mathcal{A}^{\alpha}_{\mathbf{k},x}$. In Fig.\,\ref{fig: vortices}(b) and (c) we plot the Berry connection and curvature, respectively. The opposite-chirality vortices at $X$ and $Y$ can be clearly identified, along with the resulting quadrupolar Berry curvature.

{\em  Antiferrochiral spin currents and Wilson-loop flux. ---} 
As a final consideration, we show that the squeezing vortices result in equilibrium spin currents that form an antiferrochiral lattice of loops in real space. To see this, consider the transverse energy of a nearest neighbor bond
\begin{align}\label{eq: hperp}
    h^\perp_{\mathbf{r}\mathbf{r}'} = \Delta_{\mathbf{r} \mathbf{r}'} S_\mathbf{r}^+S^-_{\mathbf{r}'}+\Delta_{\mathbf{r} \mathbf{r}'}^* S_{\mathbf{r}'}^+S^-_{\mathbf{r}},
\end{align}
where $S^\pm_{\mathbf{r}} = S^x_\mathbf{r} \pm i S^y_\mathbf{r}$, and $\Delta_{\mathbf{r}\mathbf{r}'} = J_1\pm iD_1$ ($J_1\mp iD_1$) when $\mathbf{r}\in \Lambda^A$ ($\mathbf{r}\in \Lambda^B$) and $\mathbf{r}'-\mathbf{r} \parallel \bm{\delta}^\pm$. The spin current on this bond may be obtained by introducing a fictitious U(1) gauge field $A_{\mathbf{r}\mathbf{r}'} = -A_{\mathbf{r}'\mathbf{r}}$ on the bond $\Delta_{\mathbf{r}\mathbf{r}'}\mapsto e^{iA_{\mathbf{r}\mathbf{r}'}}\Delta_{\mathbf{r}\mathbf{r}'}$ and differentiating accordingly:
\begin{align}\label{eq: jrrp}
    j_{\mathbf{r}\mathbf{r}'} = -\frac{\partial h^\perp_{\mathbf{r}\mathbf{r}'}}{\partial A_{\mathbf{r}\mathbf{r}'}}\Big|_{A_{\mathbf{r}\mathbf{r}'} = 0} = -i\left(\Delta_{\mathbf{r} \mathbf{r}'} S_\mathbf{r}^+S^-_{\mathbf{r}'}-\Delta_{\mathbf{r} \mathbf{r}'}^* S_{\mathbf{r}'}^+S^-_{\mathbf{r}}\right).
\end{align}
Now, consider a $\ce{Se}$-centered plaquette with the base $A$-site at position $\mathbf{r}$ (see Fig.\,\ref{fig: lattice}). If $\mathbf{r}' = \mathbf{r}+\bm{\delta}^+$, then, by $C_{4z}$-symmetry (about the plaquette center), we have that $\braket{j_{\mathbf{r},\mathbf{r}'}} = \braket{j_{\mathbf{r}',\mathbf{r}+\hat{\mathbf{y}}}}$, and so on in a counterclockwise fashion around the plaquette. Similarly, translation invariance guarantees $\braket{j_{\mathbf{r}\mathbf{r}'}} = \braket{j_{\mathbf{r}+\hat{\mathbf{y}},\mathbf{r}'+\hat{\mathbf{y}}}}$, implying opposite chirality of the adjacent $\ce{O}$-centered plaquette whenever $\braket{j_{\mathbf{r}\mathbf{r}'}}\neq 0$. We find
\begin{align}\label{eq: jexp}
    \braket{j_{\mathbf{r}\mathbf{r}'}} = -8SJ_1D_1\sum_{\mathbf{k}}\frac{\cos(k_x)+\cos(k_y)}{\omega^\alpha_\mathbf{k}+\omega^\beta_{-\mathbf{k}}} 
\end{align}
which is nonzero whenever $J_1\neq 0$, $D_1\neq 0$, and $JJ_+ \neq J_1^2-D_1^2$ \cite{SM}.  A related equilibrium current was recently found in a non-centrosymmetric collinear AFM \cite{MagEqCurrentDMI_Zyuzin_arXiv2026}, arising from a DMI-induced nonreciprocal pairing (in our notation, $\Delta_{\mathbf{k}} \neq \Delta_{-\mathbf{k}}$) and forming a net, uniform flow. In our centrosymmetric altermagnet, $\Delta_{\mathbf{k}} = \Delta_{-\mathbf{k}}$, so that the net current vanishes. Instead, staggered, counter-circulating loops of nonzero spin current originate in the vortex structure of $\Delta_{\mathbf{k}}$.

The origin of these currents can be further understood from the perspective of enclosed fluxes. Consider a local U(1) transformation $\mathbf{S}_\mathbf{r}\mapsto R_z(\varphi_\mathbf{r})\mathbf{S}_\mathbf{r}$ so that $\Delta_{\mathbf{r}\mathbf{r}'}\mapsto e^{i(\varphi_{\mathbf{r}}-\varphi_{\mathbf{r}'})}\Delta_{\mathbf{r}\mathbf{r}'}$. It follows that the Wilson loop
\begin{align}
    W_{\mathcal{C}} = \prod_{\braket{\mathbf{r}\mathbf{r}'}\in\mathcal{C}}\Delta_{\mathbf{r}\mathbf{r}'},
\end{align}
with $\mathcal{C}$ any closed path on the lattice, is invariant under arbitrary local U(1) transformations. Writing $\Delta_{\mathbf{r}\mathbf{r}'} = |\Delta_{\mathbf{r}\mathbf{r}'}|e^{i\phi_{\mathbf{r}\mathbf{r}'}}$, it follows that the ground state energy cannot depend on $\phi_{\mathbf{r}\mathbf{r}'}$ individually, but rather the gauge-irremovable net flux
\begin{align}
\Phi_{\mathcal{C}}=\sum_{\braket{\mathbf{r}\mathbf{r}'}\in\mathcal{C}}\phi_{\mathbf{r}\mathbf{r}'}.
\end{align}
For $\mathcal{C}$ a counterclockwise-oriented \ce{Se} (\ce{O})-centered plaquette, we have $\Phi^{\ce{Se} (\ce{O})}_{\diamond} = (-)4\phi$, which is nontrivial whenever $\phi \neq n\pi/2$, $n\in\mathbb{Z}$. If $\phi = n\pi/2$, then one of $J_1$ or $D_1$ vanishes and $\braket{j_{\mathbf{r}\mathbf{r}'}} = 0$. Moreover, for the \ce{Se} (\ce{O})-centered plaquette, $\braket{j_{\mathbf{r}\mathbf{r}'}} = -\partial\braket{h^\perp_{\mathbf{r}\mathbf{r}'}}/\partial \Phi^{\ce{Se} (\ce{O})}_{\diamond}$ directly. That is, the enclosed fluxes are produced by an effective U(1) gauge potential. In terms of magnons, $\Delta_{\mathbf{r}\mathbf{r}'}$ maps onto a two-mode squeezing amplitude, in which case $\Phi \neq 2n\pi$ serves as an obstruction to choosing a global squeezing axis around a plaquette.  

{\em Discussion.---} In this work, we have shown that a locally staggered, symmetry-allowed DMI endows the collinear ground state of a centrosymmetric $d$-wave altermagnet with a frustrated, current-carrying quantum vacuum. The complex two-magnon squeezing amplitude vanishes at the high-symmetry points $X$ and $Y$, about which its phase winds with opposite senses. This gauge-irremovable vortex-antivortex pair cannot be undone by any smooth frame transformation, imprints itself on the anomalous ground-state correlations, generates a quadrupolar Berry curvature, and, in real space, produces an antiferrochiral lattice of counter-circulating equilibrium spin currents. Crucially, none of this structure is visible in the static spin configuration, which remains perfectly collinear: the frustration resides entirely in the squeezed quantum vacuum.

A direct experimental target is the antisymmetric part of the dynamical spin-correlation tensor, whose momentum dependence should reverse handedness between $X$ and $Y$. 
Separately, boundaries, defects, domain walls, or any distortion or interface that makes the two plaquette environments inequivalent can locally lift the loop compensation, producing net current flow and thus observable local torque, spin edge accumulation, polarization, or secondary magnetization. Such effects may offer a sample-dependent contribution to the weak ferromagnetic component reported in LOMS, whose microscopic origin remains unresolved and for which no static canting has yet been established \cite{LOMSNi2010,LOMSKoo2012,LOMSKen2025,LOMSAsai2026,LOMSGarciaGassull2026,LOMSzhang2026}. Existence of DMI in these materials may be established by a measurement of the spin Nernst effect \cite{cui2023efficient,syljuasen2025quantum}.

Future work should address how interactions beyond linear spin-wave theory renormalize and potentially soften the $X$ and $Y$ modes, possibly driving a 
stripe-order instability. A central question is which phases emerge beyond this instability—including multicomponent, stripe, or chiral condensates with spin-superfluid character—and how they couple to itinerant electronic degrees of freedom. Such phases would provide an insulating magnetic counterpart to phenomena that remain elusive in strongly correlated electronic systems.

{\em Acknowledgements.---}
The authors would like to thank Violet Williams and Rule Yi for fruitful discussions. B.F. acknowledges support  from the National Science Foundation under Grant No. NSF DMR-2144086.

\appendix

\end{document}